**Picosecond UV Single Photon Detectors with Lateral Drift Field: Concept and Technologies**

**M. Yakimov[1], S. Oktyabrsky[1] and P. Murat[2*]**

[a] *SUNY College of Nanoscale Science and Engineering, Albany, NY 12203, USA*
[b] *Fermi National Accelerator Laboratory, Batavia, IL, 60510, USA*

Group III-V semiconductor materials are being considered as a Si replacement for advanced logic devices for quite some time. Advances in III-V processing technologies, such as interface and surface passivation, large area deep submicron lithography with high-aspect ratio etching primarily driven by the MOSFET development can also be used for other applications. In this paper we will focus on photodetectors with the drift field parallel to the surface. We compare the proposed concept to the state-of-the-art Si-based technology and discuss requirements which need to be satisfied for such detectors to be used in a single photon counting mode in blue and ultraviolet spectral region with about 10 ps photon timing resolution essential for numerous applications ranging from high-energy physics to medical imaging.

---

\* *Contact author, email: murat@fnal.gov*

## 1. Introduction

After over 40 years of successful development of various kinds of semiconductor photodetectors (PDs), it is quite surprising that there are still many applications utilizing vacuum tubes for light detection. It is true that for fast detectors electron transport in vacuum has a fundamental advantage over the transport in solids: lack of scattering that results in higher electron velocities and allows for reduction of internal capacitance of the vacuum device. On the other hand, recent progress in scaling of semiconductor devices and introduction of novel, other than silicon, materials into integrated circuit technologies provide new possibilities to fill the gaps between semiconductor and vacuum PDs. One of the gaps is a high-speed PD with large sensitive area and high single photoelectron sensitivity in the UV region that is required in various areas of scientific research and industry.

A number of such applications come from particle physics and astrophysics. Many modern dark matter and neutrino detectors use liquidified noble gases, LAr or LXe, which emit scintillating light at 128 nm and 175 nm, respectively. In absence of photodetectors sensitive at these wavelengths, the most common technique of detecting the emitted UV light is relying on the wavelength shifting [1]. At LHC, to identify individual proton-proton interactions and separate interactions of interest from the background in high-luminosity mode, the experiments will need timing detectors with picosecond-level resolution [2]. Detection of UV photons in the range of 90-300 nm also sheds light on many important processes in astrophysics [3]. Another application area which could greatly benefit from very fast photodetectors is the positron emission tomography (PET). PET scanners with timing resolution of the order of 10 ps could have significantly improved sensitivity leading to a breakthrough in the early diagnostics of cancer diseases. Progress in PET, however, also depends on the development of new fast scintillators which often, like in case of $BaF_2$ or $LaBr_3$, emit significant fraction of light in the UV region and require UV-sensitive photodetectors to read the signals out [4, 5].

Solid state PDs can provide separately ultra-high speed response and single photoelectron sensitivity [6, 7]. The best timing response is typically achieved in group III-V semiconductor p-i-n or metal-semiconductor-metal (MSM) PDs, while single photoelectron response requires avalanche multiplication region and low dark count rates that are best realized in Si p-i-n diodes and more recently were demonstrated in III-V diodes [8, 9]. State of the art Si solid state photomultipliers (Si avalanche diodes with on-chip signal processing circuits) demonstrate single-photon sensitivity with ~100 ps range timing response and dark count of about $10^6$ $s^{-1}$ $cm^{-2}$. Their spectral sensitivity extends down to 300 nm, covering a large part of high-speed scintillator emission spectrum.

Combining fast response and single-electron sensitivity with large sensitive area and high UV quantum efficiency is even more challenging. Semiconductor surface has always been considered a difficult part of a PD due to surface generation/recombination, surface leakage and surface traps related instabilities. As a result, detectors with low dark current and high quantun efficiency are typically made using bulk absorption in a vertical junction. Surface passivation and guarding are used when high lateral electric fields are employed in avalanche junctions. In the following, we discuss how the existing challenges can be addressed by using different detector materials along with the specific design rules. In particular, close-to-surface UV absorption with lateral field detector configuration allows for separate optimization of absorption and multiplication cross-section areas to minimize capacitance and dark current.

## 2. Detector geometry

The fastest semiconductor PDs demonstrated so far utilize lateral metal-semiconductor-metal (MSM) architecture (Fig. 1a). It has been proven to be reasonably easy to achieve rise-time below 10 ps and even below 1ps if materials with short recombination times are used [10-13]. However, this comes at a cost of reduced PD efficiency. Significant difference between MSM detectors and more commonly used p-i-n PDs is the difference in the drift field direction: the drift field is parallel to the detector surface in MSM's and orthogonal to the surface in p-i-n PDs. Both designs have their respective advantages and disadvantages in terms of temporal resolution and spectral sensitivity. Capacitance and associated RC delay, transit time of photocarriers and the absorption length are the major parameters affecting the timing and efficiency of the PD.

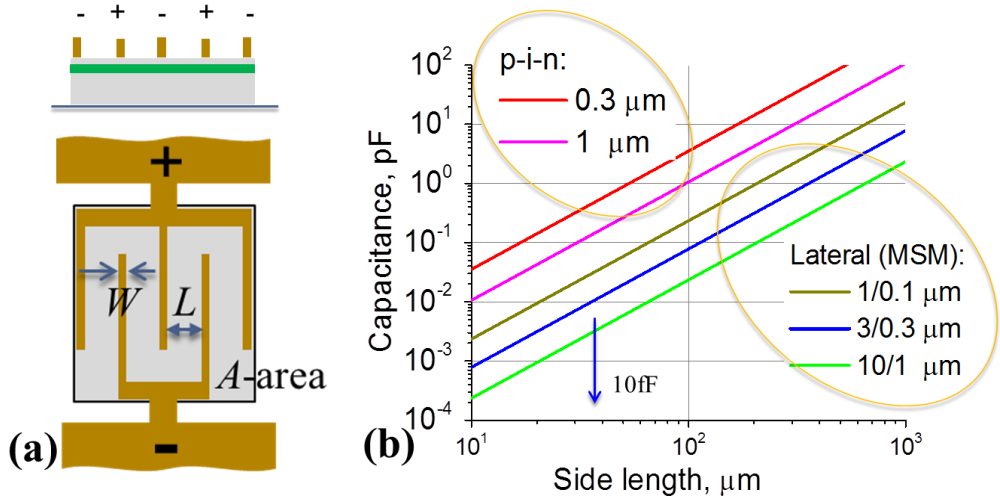

Fig. 1. (a) Top view and cross-section schematics of a PD with lateral field (e.g. metal-semiconductor-metal- MSM PD); and (b) comparison of capacitances of square devices with vertical (p-i-n) and lateral field. The MSM shows 5x reduction of device capacitance over the p-i- diode with the same drift length.

### 2.1. Detector capacitance

For p-i-n type photodetectors, the entire photodetector area can be considered as a parallel plate capacitor with plate separation equal approximately to the thickness of an i-type semiconductor. As a result, a large area PD has relatively high capacitance, which complicates design of fast detectors. Fig. 1(b) compares intrinsic capacitances of p-i-n and MSM devices with square sensitive area as a function of the side length. The capacitance of lateral field structures as in Fig. 1(a) can be easily calculated analytically [14]:

$$C = \frac{K(k)}{K\left(\sqrt{1-k^2}\right)} \frac{\varepsilon_0 (\varepsilon+1) A}{4(L+W)} \quad \text{with} \quad k = \tan^2 \frac{\pi W}{4(L+W)} \quad \text{and} \quad K(k) = \int_0^{\pi/2} \frac{d\varphi}{\sqrt{1-k^2 \sin^2 \varphi}}, \quad (1)$$

and is about 5 times less than that of p-i-n diodes with the same drift length (Fig. 1b). One of the drawbacks of the lateral field devices is that part of the sensitive area is shaded with the contacts. In the graphs presented in Fig. 1, the contact area is fixed at 10% of the entire device area. It should be noted that most of the demonstrated ultrafast MSM devices have very moderate 50-

70% sensitive area limited by the need for long low-resistance metal lines. However, novel IC interconnect technologies using low-resistive Cu in high aspect ratio wires make 1 kΩ (e.g. 0.1x0.1x30 μm$^3$) wires feasible. Using them would allow to increase significantly the sensitive area of the device without affecting its intrinsic RC delay.

2.2. <u>Transit drift time</u>

Intrinsic time jitter of a PD is limited by the fluctuations transit time of faster photocarriers (typically electrons) through the depletion region of the diode, which is determined by carrier drift velocity.
Drift velocity saturates in high electric fields due to optical phonon scattering and places the limit for intrinsic delay time. The delay may be reduced by thinning down the drift (depletion) region but that increases the device capacitance. Therefore, to keep the capacitance low one needs to further limit the sensitive area, In addition, thin sensitive layer may reduce the absorption efficiency, although it is typically not the case in the blue/UV spectral region. The interplay between the drift length and capacitance determines the device area and depletion width for the required PD response time. Table 1 shows saturation properties and corresponding device parameters for PDs with 10 ps rise-time built using different materials. One of the most obvious advantages of III-V materials with low electron effective masses is the much higher saturation drift velocity achieved at lower fields. Compared to Si p-i-n PDs, a 3x increase in the electron drift velocity combined with 5x reduction of capacitance allows to improve the speed of III-V PDs with lateral field by a factor of ~15. Alternatively, one could increase the PD sensitive area by the same factor, keeping the timing response unchanged.

Table 1. Saturation field ($E_{sat}$), velocity ($v_{sat}$), length for 10ps drift and corresponding operational voltage ($V_{sat}$) for different PD materials.

| Material | $E_{sat}$, kV/cm | $V_{sat}$, x10$^5$ m/s | Spacing for 10 ps, μm | $V_{sat}$, V |
|---|---|---|---|---|
| **Si** | 20 | 0.8 | 0.8 | 1.6 |
| **GaN** | 150 | 2.4 | 2.4 | 36 |
| **GaAs** | 3.5 | 2 | 2 | 0.7 |
| **InP** | 12 | 2.5 | 2.5 | 3 |
| **InGaAs/ InP** | 5 | 3 | 3 | 1.5 |

Lateral field PDs with lithographically defined wires have demonstrated the shortest drift-limited rise-times as shown in Fig. 2[10-13] [15, 16]. It is worth noting that the transit time slightly increases with the absorption length because photoelectrons travel deeper in the PD structure [17]. This effect vanishes if the absorption length is less than the spacing between the electrodes.

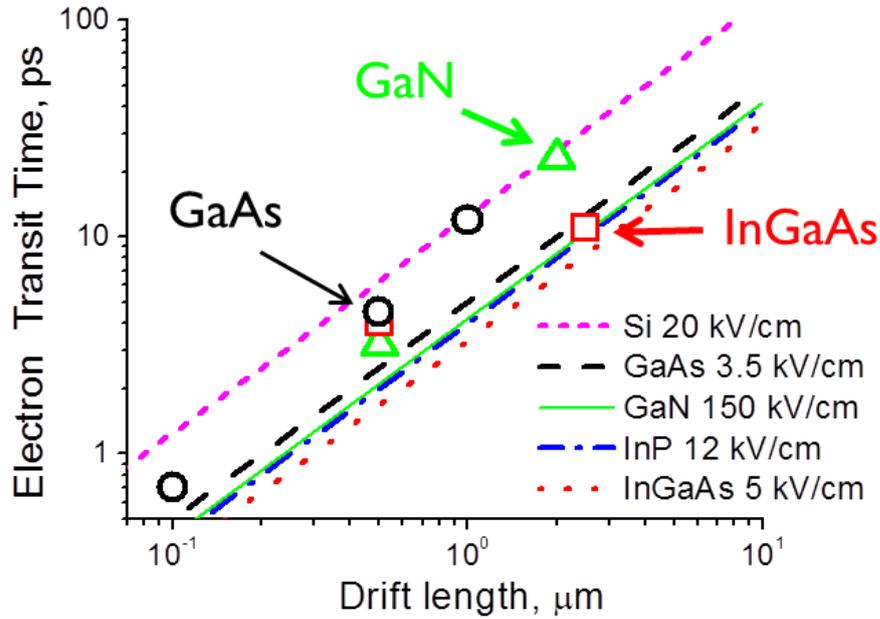

Fig. 2. Saturation velocity-limited electron drift transit time in various materials. Symbols indicate the demonstrated MSM rise-times [10-13].

2.3. Spectral response

In order to create a p-i-n PD, a highly-doped conductive semiconductor layer needs to be formed along the surface. That creates no issues for visible light detectors since light penetrates reasonably deep into semiconductor. However, the short UV absorption length in most common materials makes absorption within the doped layer a significant problem. In 300-400 nm interval, the absorption coefficient in most of the III-V's is about $7\times10^5$ cm$^{-1}$, and photocarriers are generated in a ~20 nm thick surface layer. Silicon has even higher absorption $>10^6$ cm$^{-1}$ for the wavelengths below 370 nm and, therefore, light is absorbed within just 10 nm from the surface.

Since a substantial number of photocarriers is generated in a highly doped contact region with low electric field, the high absorption value significantly reduces quantum efficiency of Si p-i-n devices in the UV region [18]. As a result, both detector efficiency and response time degrade due to recombination and slower carrier diffusion in the highly-doped areas. In an effort to reduce the thickness of the highly-doped contact layer, the inversion-layer and delta-doped structures were introduced in so-called UV-enhanced PDs [19]. The latter design has shown a very impressive UV efficiency up to 60% at the wavelengths down to 125 nm in small 4x4 μm$^2$ CCD pixels [20]. These approaches relying on reduction of the contact layer thickness inevitably increase the lateral resistance of this layer connected in series and add up to the RC delay detrimental for high-speed applications. In fact, a significant slow diffusion component of the temporal response and relatively high attainable sheet resistivities close to ~10 kΩ/sq. make the design of ultrafast UV-enhanced PDs a real challenge.

Total elimination of highly-doped surface contact layer is possible in lateral field (e.g. MSM) devices. The lateral field geometry becomes favorable when material/wavelength combination with high absorption coefficients is employed, such as III-V semiconductors in blue and UV spectral regions. The lateral field that exists on the very surface of the structure (as well as above the surface) makes collection and drift of photocarriers efficient and fast.

Table 1 shows the operating parameters for several III-V semiconductor devices with the lateral field. The required operating voltages are quite low in all the cases, except for GaN. In addition, GaN being a wide-bandgap material has significantly higher absorption length which could complicate design of a detector with a thin surface quantum well (QW) and could require a different geometry and higher voltages.

One of the issues with the III-V materials is the high surface recombination rate which reduces the response time but also degrades quantum efficiency. Modern surface and interface passivation techniques developed for novel III-V MOSFETs [21] significantly reduce the interface trap density and recombination rates so that the photocarrier drift transit time is not affected. However, as described below, the surface/interface might still be a source of carrier generation that affects the device dark current.

### 3. Dark Current and Dark Count Rate

Noise is a critical factor in any detector applications, but the need of high single photon quantum efficiency makes the low noise requirement especially demanding. Despite the wider bandgap of GaAs compared to Si, the dark current density in III-V photodetectors is higher than in Si PDs. The reasons for the increased dark current appear to be technology-related, with the generation (commonly called generation/recombination – GR) current within GaAs depletion region being a primary source of dark current.

At reverse bias voltage saturating the drift velocity, there are two processes contributing to the PD dark current: injection of carriers from the contacts and carrier generation in the photosensitive (depletion) region of the semiconductor. An MSM detector fabricated with two identical metal-semiconductor junctions would have at least one contact with the barrier height of $E_g/2$ or less ($E_g$ is the semiconductor bandgap). As a result, the dark current is mostly determined by thermionic injection of carriers over the lowest MS barrier $\phi_B$:

$$I_{thermionic} \sim T^2 \exp\left(-\frac{q\phi_B}{kT}\right) \qquad (2)$$

In semiconductor PDs with p-n or p-i-n junctions, the dark current is dominated by the generation in the depletion region, which include the bulk Shockley-Hall-Read (SHR) generation and generation through surface states. The SHR dark current can be calculated as

$$I_{SHR} \approx qV_{depl}\frac{n_i}{2\tau_o}, \qquad (3)$$

where $V_{depl}$ is the volume of high electric field (depletion) region, $n_i$ – intrinsic carrier concentration, and $\tau_0$ – carrier recombination lifetime (assuming equal lifetime for electrons and holes). The surface component of the generation current depends on the PD geometry. For example, in a p-i-n PD it is determined by the perimeter area with high electric field and surface trap density.

Figure 3 shows a simulated dependence of dark current density vs. operating voltage in 1μm long/thick MSM or p-i-n PDs. The effect of thermionic emission on the barrier height of a semiconductor contact and generation/recombination is clearly seen. For identical metal-semiconductor junctions of an MSM detector, at least one of the barriers to GaAs with $E_g$=1.42eV is ~0.7 eV or below, resulting in the injection current higher than ~30 μA/cm² at room temperature [22, 23]. To reduce the thermionic current, a metal contact to a wider bandgap material such as AlGaAs or InAlGaP is required [24-27]. A built-in heterojunction barrier close to the Schottky contact, as shown in the inset of Fig. 3, can further suppress the current by 3-4 orders of magnitude to ~10 nA/cm$^2$ range. So far the lowest dark current obtained in MSM PDs with symmetrical contacts is about 200 nA/cm$^2$ (dark count rate ~10$^4$ e/s-μm$^2$) at room temperature [27] that is likely related to tunneling through the defects in Al-containing capping layers. Further lowering the dark current can be done by utilizing separate higher barrier metals for cathode and anode. When the thermionic current is low enough, the generation/recombination current component in Eq.(3) starts dominating as indicated in the Fig. 3 for AlGaAs with and without GR contribution.

Since in p-i-n diodes the barriers for minority carrier injection are close to the bandgap energy, the reverse current is mostly due to the GR. Interestingly, though the intrinsic carrier density in a wider bandgap GaAs is four orders of magnitude less than in Si, the GR current in GaAs p-i-n PDs is 1-2 orders higher (Table II). In commercially available GaAs p-i-n PDs, the dark current is typically ~10-30 nA/cm$^2$ [28] that is way higher than 0.1 nA/cm$^2$ in Si devices [29]. This difference is attributed to significantly higher non-radiative defect density in conventional MOCVD- or MBE-grown binary group III-V materials and is even worse in ternary III-V alloys. In general, an improvement of bulk material quality achieved by using, for example, liquid-phase epitaxy allows to reduce the reverse currents in GaAs diodes down to the level of ~50 pA/cm$^2$ [30]. However, it is unlikely that any technology other than MOCVD or MBE can be used in fabrication of real devices as these technologies have numerous advantages in growing the thin film heterostructures.

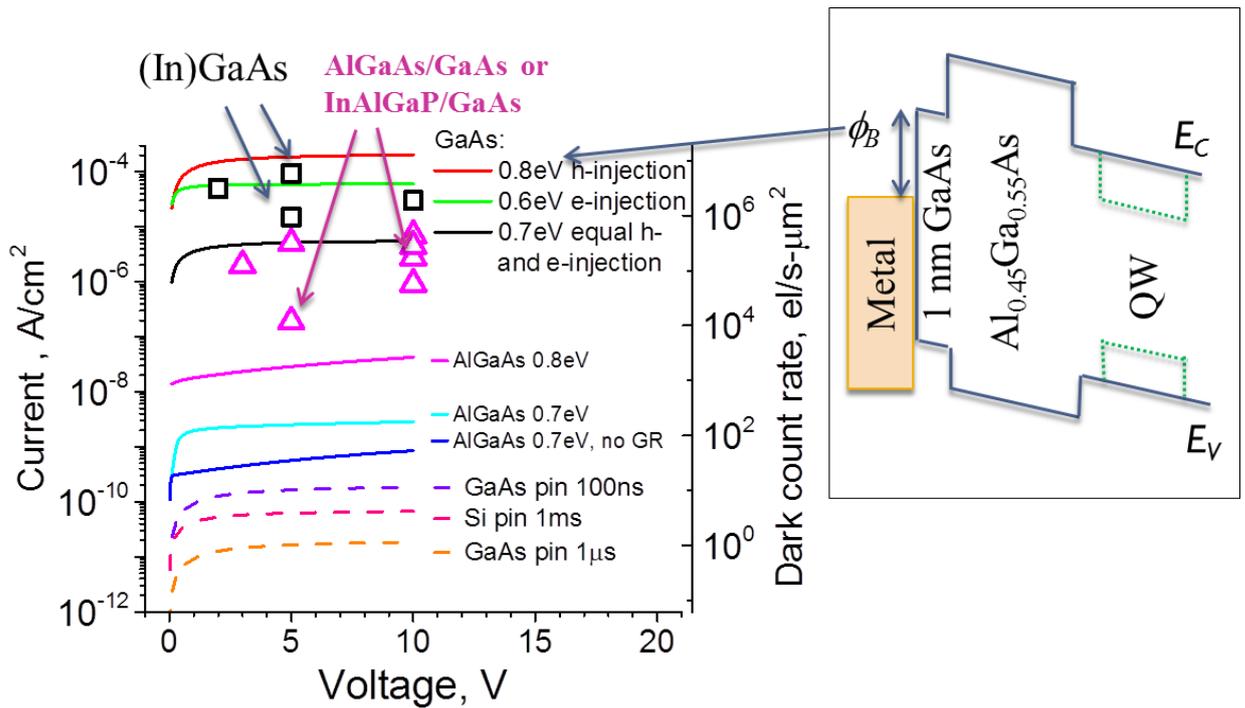

Fig. 3. Simulated room temperature dark current and dark count rate for different barriers to i-region of an MSM PD and different recombination lifetime in p-i-n PD showing the effect of thermionic emission and generation/recombination. The device length is kept at 1μm. The inset shows a band diagram of the simulated Schottky contacts with wider-bandgap material to reduce injection and with a quantum well conducting channel in a lateral field PD to reduce generation/recombination volume. Symbols indicate experimental dark currents of MSM PDs with different barriers from Ref. [22-27].

Another way to reduce the GR current detectors with lateral field is to reduce the volume of the depletion region $V_{depl}$ in Eq.(2). The Fig. 3 inset shows the cross-section bandstructure of a lateral-field PD with a quantum well inserted for the photocarrier transport. If the barrier bandgap is wide enough to provide low carrier density, the $V_{depl}$ will be given by the QW volume or more accurately by the volume of the quantum confined envelope wavefunctions of carriers in the QW.

Table II. Estimated material parameters and GR dark current in 0.5 μm p-i-n diodes.

| Material | $n_i$, cm$^{-3}$ | $\tau_o$, s | $J_{GR}$, A/cm$^2$ |
|---|---|---|---|
| Si | $1.5 \times 10^{10}$ | $10^{-3}$ | $6 \times 10^{-11}$ |
| GaAs | $1.8 \times 10^{6}$ | $10^{-8}$ | $7 \times 10^{-10}$ |
| Al$_{0.3}$Ga$_{0.7}$As | $1.7 \times 10^{3}$ | $10^{-10}$ | $7 \times 10^{-11}$ |

The GR current in lateral-field PD will be also significantly affected by the surface GR since the surface area in these detectors is relatively larger than in p-i-n PDs. The structure design should include wide-bandgap top barrier to "isolate" the photocurrent in the QW from the surface. In addition, to reduce the GR rate down to bulk values novel methods of GaAs surface passivation developed for III-V MOSFETs have to be used as described later (Section 6).

Concluding this section, it is worth summarizing the design rules to achieve low dark current in the lateral field III-V PDs:
− Proper design of the contact structure, using either a lateral p-i-n structure or a wide-bandgap barrier material instead of simple metal Schottky contacts;
− Minimization of the area of the contact;
− Reduction of the volume of material with the highest GR current, for example, using the GaAs quantum well for carrier transport with wide-bandgap barriers as in the inset of Fig. 3;
− Reduction of the surface/interface GR current by placing the high bandgap barrier on the surface and surface passivation to reduce trap interface trap density.

## 4. Internal Amplification and Photon Timing

Owing to a relatively high capacitance and noise of amplifiers, single photon or more accurately single photoelectron detection requires internal amplification. Over the years, an avalanche amplification has become a standard for single photon semiconductor detectors [31]. The avalanche build-up and relaxation are relatively slow processes; therefore, it is important to evaluate whether they allow for the ps-range single photon timing. Similarly to photomultiplier tubes, in avalanche PDs (APD) the photon timing is achieved by discriminating the leading edge of the single photoelectron pulse just above the noise level. In this case, the long avalanche relaxation does not contribute to the photon timing jitter, but rather increases the detector dead time.

In many solid-state photon counters, the APD is used in a Geiger mode, when the diode current is zero until a photocarrier reaches the multiplication region and triggers a large avalanche current. The diode is further quenched by an external passive or active circuit. Although the multiplication factor in Geiger mode is very high (up to $10^6$), the timing of the photon arrival is determined by a fast discriminator which senses the leading edge and is set low to optimize time resolution. Therefore, the time resolution is coupled to an early stage of the avalanche development, even though the APD may operate at very high gain.

The above considerations illustrate the difference between two main single photon timing parameters: pulse delay and time resolution which correspond to the circuit risetime and jitter, respectively. The single photoelectron pulse risetime will be mostly affected by the drift transit time in the absorption region and multiplication process. On the other hand, the geometrical uncertainty of photon absorption and stochastic nature of the avalanche process will contribute to the jitter/resolution. Depending on a specific process, the time resolution can be 3-10 times shorter than pulse risetime. This allows for very fast single photon timing (time resolution) of 20-30 ps in Si APDs [32, 33] and will likely allow ~1 ps resolution in the III-V PDs with a 10 ps pulse risetime.

*4.1 Avalanche build-up*

The rise time of an APD photocurrent is determined by multiple transport of carriers through the avalanche region as shown in Fig. 4(a). The avalanche needs time to build-up with a contribution from both slow holes and fast electrons. Quite simplistically, it can be assumed that the rise time is determined by the saturation drift velocity of the carriers through the avalanche region. The detailed Monte-Carlo simulations show that the actual carrier velocity may be significantly higher than the saturation velocity [34], and the resultant response might be somewhat faster [35]. At high frequencies, the build-up time (Fig. 4b) increases almost linearly with the multiplication factor, and this slope is characterized by the gain-bandwidth product (GBP). Reduction of the avalanche region thickness increases the GBP, but at the thickness below ~50-100 nm, the electric field needed for carriers to attain kinetic energy sufficient for impact ionization becomes too high. The highest demonstrated avalanche GBPs are about 300-400 GHz in various kinds of APDs [36, 37].

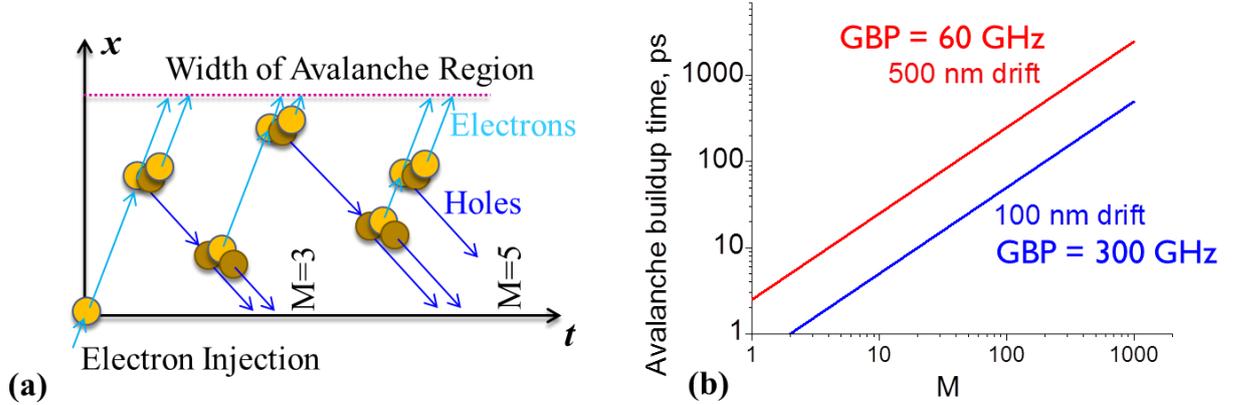

Fig. 4. (a) Avalanche multiplication of photocurrent illustrating build-up time increase at high multiplication factor M. (b) Avalanche build-up time in a separate absorption-multiplication (SAM) GaAs APD with two multiplication region thicknesses.

*4.2. Threshold multiplication factor*

Rough estimation of the circuit requirements for a ps-range APD uses the equivalent circuit shown in Fig.5 that include APD noise current source $i_{APD}$, total capacitance $C$, and equivalent noise input resitance $R_{eq}$ with its noise source $i_{amp}$. The pulse amplitude due to a single photoelectron can be estimated as

$$V_{ph} = \frac{I_{ph}\tau}{C} \xrightarrow{single\ ph.e.} \frac{Me}{C} \qquad (4)$$

where the total number of electrons, multiplied by the factor $M = GBP \cdot 2\pi\tau$, gives the charge on the capacitor $C$. Although the capacitor discharge may reduce this amplitude, in typical circuits the discharge time is much longer than the rise time $\tau$. Fig. 5(b) illustrates, for different capacitance values, the increase of the signal amplitude with the rise time that limits the multiplication factor, and as the circuit capacitance is reduced. The signal amplitude should exceed the noise level of the loading resistor and the amplifier. Considering a high electron mobility transistor (HEMT) as an amplifier with the equivalent Johnson noise resistor of 100 Ω (corresponding input noise 2.3 nV/√Hz), the noise voltage amplitude in the bandwidth $\Delta f = 1/2\pi\tau$ is

$$v_{noise}^2 = 4\pi k T R_{eq} \Delta f \quad , \tag{5}$$

and the dependence of the noise amplitude on rise time is plotted in Fig. 5(b). The graph gives some useful target parameters for the device design. For example, for the target rise time of 10 ps (threshold M=20 from the Fig. 4b) the amplifier noise determines the maximum capacitance of 20 fF to achieve unity signal-to-noise ratio.

These simple arguments leave quite limited room for device design variations. Very low capacitance of ~10 fF requires the avalanche region of the device with area as small as ~10 μm$^2$ and 100 nm thick for high GBP. It also likely requires a monolithically integrated FET amplifier since the chip-level integration schemes typically have capacitance well above 0.1pF per bond.

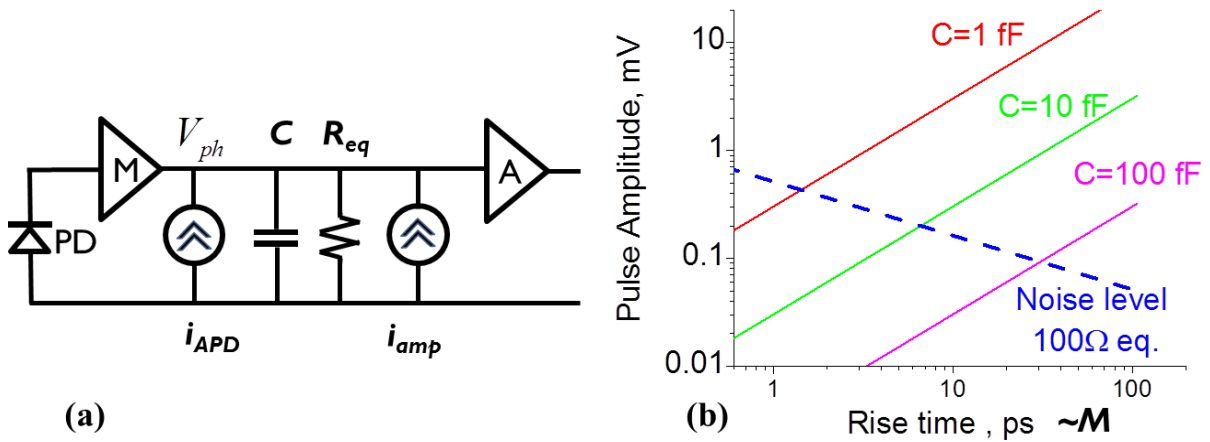

Fig. 5. (a) Equivalent circuit of APD with preamplifier. (b) Dependence of a signal pulse amplitude on pulse rise time in an APD with GBP= 300 GHz and load capacitance. HEMT noise is shown with a dashed line.

The choice of material and design of the avalanche region is dictated by conventional requirements for the lowest excess noise and stability against the microplasma discharge. The excess noise is predicted by the McIntyre theory that projects the lowest noise factor for materials with very different impact ionization coefficients for electron and holes. Their ratio, called k-ratio, is relatively low ~0.2 for Si and is much closer to unity for GaAs (~0.8), therefore, Si APDs show lower noise. Moreover, in thiner avalanche regions that require higher electric field the ionization coefficients of electrons and holes are both increasing and approaching each other, and the k-ratio is approaching unity. However, recent experimental results show that as the avalanche region becomes thinner the excess noise is decreasing instead of increasing [38]. Even more promising are the continuing efforts to engineer impact ionization region using multilayer III-V structures [39-41]. This approach results in an increase of the kinetic energy of injected electrons and better localization of the impact-ionization events; as a result a very low excess noise factor of 3.3 at M=16 in a thin 260 nm wide bandgap $Al_{0.6}Ga_{0.4}As$ multiplication region was demonstrated [42].

## 5. Device design

Implementation of the approaches described above into a device structure leads to a design shown in Fig. 6. The structure epitaxially grown on semi-insulating GaAs substrate consists of two distinct regions: the lateral absorption region and the vertical avalanche region. The absorption region utilizes a depleted GaAs or InGaAs quantum well (QW) placed between variable bandgap $Al_xGa_{1-x}As$ barrier layers. The Al content $x$ is changing (Fig. 6 inset) in order to provide a built-in electric field for fast and efficient collection of photocarriers into the QW. The variable bandgap AlGaAs barrier thickness ~100 nm is chosen to ensure the complete absorption of light in the selected spectral range. The surface of the absorption region is passivated, e.g. by an atomic layer deposited $Al_2O_3$ layer on top of a few monolayers-thick GaAs to prevent oxidation of the AlGaAs barrier. Additionally, the oxide window can be used as an anti-reflection coating.

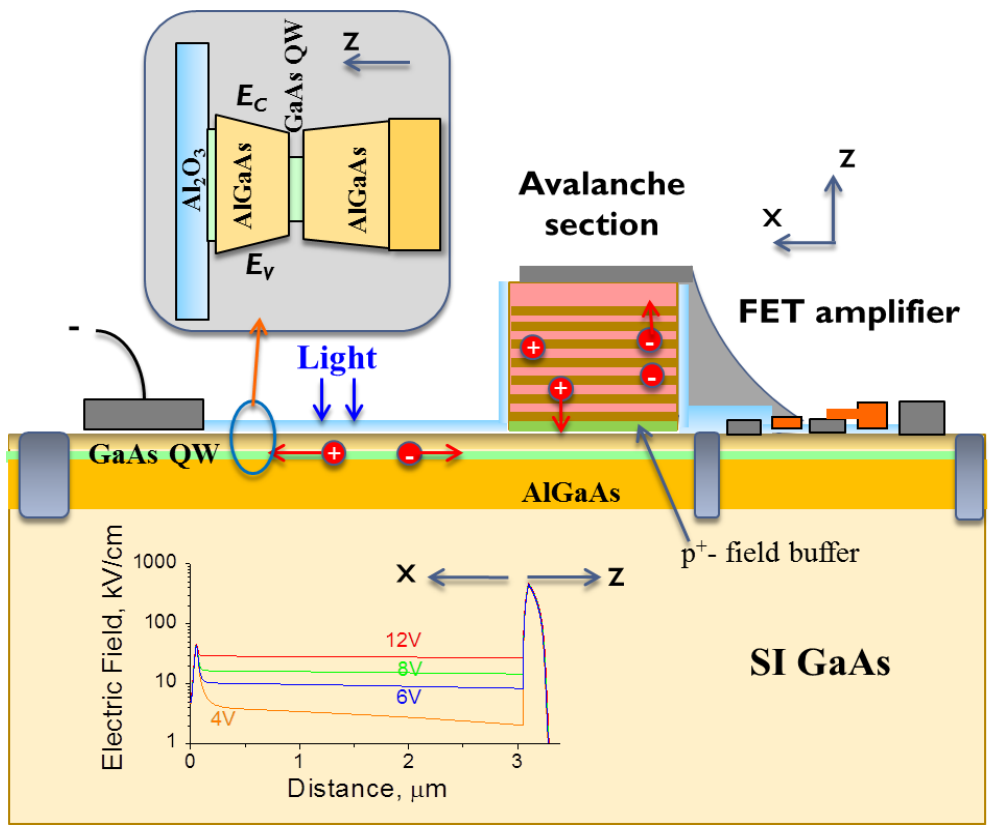

Fig.6. Design concept of ps-range single photon detector consisting of separate lateral field photodiode, vertical avalanche region and integrated FET. The inset shows the band structure of the top PD layers optimized for UV/blue spectral range. The graph illustrates the electric field in the device at various biases.

The left contact in Fig. 6 could utilize either epitaxial $p^+$-GaAs junction or Schottky barrier with a high work-function metal that create minimal intermetallics with GaAs, such as Pt or Ti. These types of contacts have large barriers for electron injection and will minimize the dark current associated with contacts. The generation/recombination current in the absorption region is minimized by a relatively small volume of the device and reduced surface recombination due to the adequate passivation layer.

Due to orthogonal fields, the absorption and the avalanche regions can have different cross-section areas. As the avalanche region should be thin (<100nm) to provide fast multiplication build-up, its cross-section should be minimized to give low capacitance. A $p^+$-field buffer separates the undoped absorption region with relatively low electric field ~10 kV/cm and avalanche region with the field >400kV/cm. The field distribution is also shown in the inset of Fig.6. The avalanche region will likely employ AlGaAs or superlattice multiplication structure as discussed above. Although there are significant technological challenges, such as reduced defect density, surface leakage and isolation, avalanche stability against microplasma discharge/breakdown, which need to be dealt with, there are no physical reasons preventing this device from operating as expected.

## 6. Device integration

As described above, one of the major limiting factors of the fast PDs is their capacitance which proportionally reduces the pulse amplitude. As a result, faster detection of weak signals requires smaller area PDs. The fastest silicon avalanche PDs of 10 μm diameter can detect photon arrival time with resolutions down to 30-40 ps or even 20 ps for 5 μm PD [32, 33].

Since many applications need large area detectors, the capacitance is reduced by partitioning of the photodetector area into array of small pixels. PDs with analog summation of signals from the individual pixels are available from Hamamatsu, RMD and others, digital circuits have been recently implemented by Philips [23, 24].These "Si photomultipliers" (SiPMTs) are becoming now a standard for single photon detectors. Although the partitioning of large area (mm and cm size) detectors is inevitable for ps-range detectors, the partitioning size requirement can be substantially relaxed in case of PDs with lateral field.

As illustrated in the Figs. 1 and 5(b), a lateral field GaAs pixel with 30x30 μm$^2$ size will have a capacitance of 10 fF, good enough to obtain a measurable 10 ps intrinsic single photoelectron response. However, to utilize this short risetime, the input and parasitic capacitances of an amplifier should be also less than 10 fF. This requirement is relatively easy to meet in monolithic integrated circuits, but before the full-scale III-V digital circuits evolve it is likely that the PD array will have to be integrated with a silicon read-out chip containing active quenching and gating circuits [7]. Most of the chip area array interconnection technologies operate with relatively large, 10's of micrometers in size, interconnects (solder bumps, bridges, and filled vias) with a capacitance exceeding 100 fF per bond. To accommodate this large parasitic capacitance, the APD output is routed to on-chip FET for further amplification. An amplifier, such as a high electron mobility transistor (HEMT) or MOSFET should be placed in a close vicinity to the PD pixel (Fig.6). Lateral field PD geometry is well-suited for monolithic integration with field-effect transistors as they both utilize similar basic technologies. In addition, the FET in Fig.6 can use the same epitaxial QW layer structure for the channel as for the PD absorber.

The integration of MSM PD with MESFETs and HEMTs has been established for a long time [6, 43, 44]. However, the advancements in the group III-V semiconductor technology have only recently resulted in high-performance MOSFETs and other 3-dimensional FETs that overcome performance of Si-based transistors, and are compatible with the design in Fig. 6.

One of the long-standing challenges in III-V FETs is the semiconductor surface passivation. The latest advances in gate oxide technologies for MOSFETs made interfaces with low interface rap densities possible. Surface passivation techniques include deposition of $Ga_2O_3(Gd_2O_3)$ [45], ultrathin amorphous silicon passivation layer [46], or atomic layer deposition od $Al_2O_3$ with self-cleaning effect of TMA (Trimethylaluminium) precursor [47]. It is important that the interfaces with low density of interface traps ($10^{11}$-$10^{12}$ $cm^{-3}eV^{-1}$) also demonstrated low surface recombination rates [48]. Combined with the surface band engineering to separate the photocarriers from the oxide interface with wide-bandgap semiconductor (Fig. 6), the passivation layer can drastically reduce the surface state density. Given that the oxides ($Al_2O_3$, $SiO_2$, $HfO_2$) are large bandgap materials, they can be used as UV/visible windows protecting detector from the environment as well as an anti-reflection coating.

Based on the previous discussion, Table III summarizes how the materials and device parameters affect the characteristics of the PD. It is worth noting that the proposed device geometry allows for the best combination of large absorption area, short absorption length essential for blue/UV spectral range, low capacitance of both absorption region (due to the lateral field) and avalanche region (due to low cross-section) if high mobility/ high saturation velocity material is used.

Table III. Effect of material parameters and device design on characteristics of a UV PD with lateral field

| PD characteristics | Material and design parameters |
|---|---|
| UV Efficiency | - Surface recombination<br>- Band structure of the absorption region<br>- Avalanche initiation probability |
| Delay/latency | - Device capacitance<br>- Drift time in absorption region<br>- Avalanche build-up |
| Jitter/ time resolution | - Avalanche build-up<br>- Drift time in absorption region<br>- Device capacitance |
| Dark current/noise count rate | - Injection from contacts<br>- Surface generation/recombination<br>- Bulk generation/recombination in absorption and avalanche regions |
| Amplitude of a single photoelectron pulse | - Device and integration capacitance<br>- Avalanche multiplication |

Lastly, an area array integration with Si electronics is required to achieve functionality similar to SiPMT photodetector, including discriminating, active quenching, gating and/or time correlation circuits. Depending upon the scale, complexity and functionality of the detector, one can envision well-established varieties of flip-chip integration with solder bumps used, for example, for bonding focal plane arrays on silicon. More recently developed 3D integration technologies with wafer bonding and interconnecting through-wafer via holes could also be used.

**Summary**

We proposed a novel design of an ultrafast single photon photodetector for UV/visible spectral range and picosecond-level timing resolution. The detector has separate absorption and amplification regions. The absorption region, covering large fraction of the total device area, has lateral electric field while the low area and low capacitance avalanche region has the field electric field normal to the substrate plane. The design rules and requirements are analyzed against demonstrated devices and technologies.